%% file: ms.tex
\documentclass[twocolumn]{aastex7}

\newcommand{\LCcite}{\hyperlink{cite.Limongi2006}{LC06}}
\newcommand{\LCCCcite}{\hyperlink{cite.Limongi2018}{LC18}}
\newcommand{\CLcite}{\hyperlink{cite.Chieffi2013}{CL13}}
\newcommand{\g}{$\gamma$}
\newcommand{\A}{$\alpha$}
\newcommand{\msun}{$\rm M_\odot$}

\usepackage[load-configurations=abbreviations]{siunitx}
\usepackage{siunitx}
\usepackage{isotope}
\usepackage{hyperref}
\usepackage{nameref}

\DeclareSIUnit{\smasses}{\ensuremath{M_{\odot}}}

\begin{document}

\graphicspath{{./}{figs/}}

\title{The role of rotation on the yields of the two $\rm \gamma-ray$ emitters \isotope[26]{Al} and \isotope[60]{Fe} ejected by massive stars}

\author[0009-0005-0324-0637]{Agnese Falla}
\affiliation{Dipartimento di Fisica, Sapienza Università di Roma, P.le A. Moro 5, Roma 00185, Italy}
\affiliation{Istituto Nazionale di Astrofisica - Osservatorio Astronomico di Roma (INAF - OAR), Via Frascati 33, I-00040, Monteporzio Catone, Italy}
\email{agnese.falla@uniroma1.it}

\author[0000-0003-0390-8770]{Lorenzo Roberti}
\affiliation{Istituto Nazionale di Fisica Nucleare - Laboratori Nazionali del Sud (INFN - LNS), Via Santa Sofia 62, Catania, Italy}
\affiliation{Istituto Nazionale di Astrofisica - Osservatorio Astronomico di Roma (INAF - OAR), Via Frascati 33, I-00040, Monteporzio Catone, Italy}
\affiliation{Konkoly Observatory, Research Centre for Astronomy and Earth Sciences, E\"otv\"os Lor\'and Research Network (ELKH), Konkoly Thege Mikl\'{o}s \'{u}t 15-17, H-1121 Budapest, Hungary}
\affiliation{CSFK, MTA Centre of Excellence, Konkoly Thege Miklós út 15-17, H-1121, Budapest, Hungary}
\email{lorenzo.roberti@inaf.it}

\author[0000-0003-0636-7834]{Marco Limongi}
\affiliation{Istituto Nazionale di Astrofisica - Osservatorio Astronomico di Roma (INAF - OAR), Via Frascati 33, I-00040, Monteporzio Catone, Italy}
\affiliation{Kavli Institute for the Physics and Mathematics of the Universe, Todai Institutes for Advanced Study, University of Tokyo, Kashiwa, 277-8583 (Kavli IPMU, WPI), Japan}
\affiliation{Istituto Nazionale di Fisica Nucleare - Sezione di Perugia (INFN), via A. Pascoli s/n, I-06125 Perugia, Italy}
\email{marco.limongi@inaf.it}

\author[0000-0002-3589-3203]{Alessandro Chieffi}
\affiliation{Istituto Nazionale di Astrofisica - Istituto di Astrofisica e Planetologia Spaziali (INAF - IAPS), Via Fosso del Cavaliere 100, I-00133, Roma, Italy}
\affiliation{Istituto Nazionale di Fisica Nucleare - Sezione di Perugia (INFN), via A. Pascoli s/n, I-06125 Perugia, Italy}
\affiliation{School of Physics and Astronomy, Monash University, VIC 3800, Australia}
\email{alessandro.chieffi@inaf.it}

\begin{abstract}

We show that the observed \isotope[60]{Fe}/\isotope[26]{Al} flux ratio provided by the SPectrometer on INTEGRAL satellite ($0.24 \pm .04$) can be reproduced only if rotation is taken into account in the computation of the stellar models. Predictions from non-rotating stellar models yield to a significantly lower ratio ($0.062$), which is incompatible with the observed value. The adopted models and the associated yields are based on a combination of models already published by \cite{Limongi2018} complemented by additional ones fully consistent with the original grid, allowing a finer resolution in the initial rotational velocity distribution.

\end{abstract}


\section{Introduction} \label{sec:intro}

The first observational evidence of the presence of \isotope[26]{Al} alive in our Galaxy was obtained by the HEAO-3 experiment (Mahoney et al. \citeyear{Mahoney1982}, \citeyear{Mahoney1984}) that confirmed the early understanding that this isotope can be synthesized by massive stars (Arnett et al. \citeyear{Arnett1977}). This, given its relatively short half-life of 1.04 Myr, leads to a steady state abundance in our Galaxy sufficient to provide a diffuse emission detectable from the Earth (Ramaty \& Lingenfelter \citeyear{Lingenfelter1977}). The latest and more extended data concerning the distribution of \isotope[26]{Al} in our galaxy have been obtained by the Compton $\gamma$-ray telescope Imaging Compton Telescope (COMPTEL) onboard of the CGRO satellite. The \isotope[26]{Al} all-sky map provided by COMPTEL (Pl\"uschke et al. \citeyear{Pluschke2001}) shows that this $\rm \gamma-ray$ emitter is primarily concentrated in the Galactic disk, with a clumpy and asymmetric distribution strongly correlated with the distribution of massive stars \citep{Knodlseder1999}. 
On the other hand, the $\rm \gamma-ray$ signal from \isotope[60]{Fe}, whose half-life is 3.75 Myr,  is too weak to allow us to derive the spatial distribution or perform an imaging analysis. Therefore, the distribution of this isotope over the Galaxy is typically constrained by fitting a parameterized geometric model. Only the SPectrometer on INTEGRAL (SPI) was able to detect a (integrated) signal from the center of the Milky Way. By comparing the emission morphology maps using different candidate source tracers for both \isotope[26]{Al} and \isotope[60]{Fe} emissions, \citeauthor{Wang2020} \citeyear{Wang2020} suggested that \isotope[60]{Fe} emission is more likely to be concentrated toward the Galactic plane. The same study provides updated values for the \isotope[60]{Fe}/\isotope[26]{Al} ratio, proposing different estimates based on different approaches to the calculation of fluxes. 
According to this study, the total estimated \isotope[60]{Fe}/\isotope[26]{Al} flux ratio is $18.4\%\pm4.2\%$, under the assumption that both \isotope[26]{Al} and \isotope[60]{Fe} follow exponential disk models.
Alternatively, the flux ratio can be derived by using a set of tracer maps. Using the best fit maps, specifically the IRAS $25\mu m$, the derived \isotope[60]{Fe}/\isotope[26]{Al} flux ratio is $24\% \pm 4\%$. 
Both values are slightly higher than previous observational results obtained until late-2000s (e.g. \citeauthor{Leising1994} \citeyear{Leising1994}, \citeauthor{Smith2004} \citeyear{Smith2004}, \citeauthor{Wang2008} \citeyear{Wang2008}).

Since the early 70s, theoretical models aimed at understanding of the origin of the observed Galactic fluxes of both \isotope[26]{Al} and \isotope[60]{Fe} \citep[see, e.g.,][]{pardo1974}, together with the primordial solar-system ratio of \isotope[26]{Al} and \isotope[27]{Al} \citep{lee1976}.
Over the years, many authors tried to reproduce the observed fluxes of both isotopes by means of stellar models in various mass intervals, as detailed in the reviews by Prantzos (\citeyear{prantzos2004}) and Diehl (\citeyear{diehl2004}).
In 2006, \citeauthor{Limongi2006} (hereafter \LCcite{}) presented a comprehensive and accurate analysis of the synthesis of both these two $\rm \gamma-ray$ emitters based on a grid of non-rotating stellar models ranging in mass between 11 and $\SI{120}{\smasses}$.
After that work,  only \citeauthor{Woosley2007} (\citeyear{Woosley2007}) computed a new extended set of stellar models that could be used to interpret the observed fluxes of both isotopes.
Roughly one decade after the \LCcite{} paper, \citeauthor{Limongi2018} \citeyear{Limongi2018} (hereafter \LCCCcite{}) and \cite{roberti2024} presented a new generation of stellar models ranging in mass between 13 and $\SI{120}{\smasses}$, in metallicity between [Fe/H]=0.3 and [Fe/H]=--3 and with initial equatorial rotation velocities of 0, 150 and 300 km/s. However, they did not analyze the yields of \isotope[26]{Al} and \isotope[60]{Fe}, and in particular the effect of rotation on the interpretation of the observed fluxes of these two $\rm \gamma-ray$ emitters. In this paper we want to show the effect of rotation on the synthesis of these two nuclei, refining the grid of solar metallicity models with new computations fully homogeneous with those published by \LCCCcite{}. 
The paper is organized as follows. Section \ref{sec:1} briefly introduces the new grid of stellar models that form the basis of our analysis, while the influence of rotation on the yields of both \isotope[26]{Al} and \isotope[60]{Fe} is discussed in Section \ref{sec:2}. 
Our theoretical predictions are eventually compared with the observed Galactic \isotope[60]{Fe}/ \isotope[26]{Al} $\gamma$-ray line flux ratio in Section \ref{sec:4}. A final conclusion summarizes our findings.

\section{The stellar models}\label{sec:1}
The presupernova evolutions of all models have been computed by means of the same version of the FRANEC (Frascati Raphson Newton Evolutionary Code) code adopted in \LCCCcite{}, with the same input physics and nuclear network (see \CLcite{} and \LCCCcite{} for all the details). Let us just remind here that the network includes 335 isotopes (from neutrons to \isotope[209]{Bi}) and 3019 nuclear reactions and that all nuclei beyond Molybdenum are considered at the local equilibrium unless they are close to the neutron magic numbers 82 and 126. In this last case all processes involving their production and destruction are explicitly taken into account. The original grid of masses is sampled with 13, 15, 20, 25, 30, 40, 60, 80 and 120 $\rm M_\odot$ and three initial equatorial rotational velocities v = 0, 150, 300 km/s. Given the coarse sampling in initial rotational velocities, we have computed additional models with v = 50, 100, 200, and 250 km/s, so that our final grid of models spans a velocity range between 0 and 300 km/s, with a step size of 50 km/s. Mass loss is included following the prescriptions of Vink et al. (\citeyear{Vink2000, vink2001}) for the blue supergiant phase, \citeauthor{dejager1988} (\citeyear{dejager1988}) for the red supergiant phase and \citeauthor{nugis2000} (\citeyear{nugis2000}) for the Wolf-Rayet phase. Also mass loss triggered by the approach to the Eddington luminosity is taken into account.

The explosion has been simulated by means of the HYPERION (HYdrodynamic Ppm Explosion with Radiation diffusION) code (see \citeauthor{limongi2020} \citeyear{limongi2020} for all the details). All explosions have been computed by requiring the ejection of $\SI{0.07}{\smasses}$ of \isotope[56]{Ni} in order to leave to any potential user the freedom to choose the preferred initial mass - remnant mass relation. However, in this specific context we followed the Recommended scenario (see \LCCCcite{}), i.e. the one in which all models with an initial mass $M\le\SI{25}{\smasses}$ eject 0.07 \msun of \isotope[56]{Ni} while the more massive ones fully collapse in the remnant, contributing to the yields only through the wind. 

In Tables \ref{tabyields_050}, \ref{tabyields_100}, \ref{tabyields_200} and \ref{tabyields_250} we provide the yields of all the isotopes for all the new models. Since, as already mentioned, the yields of stars with $M\leq \SI{25}{\smasses}$ are those with both contribution, i.e. the wind and the explosion, for the sake of completeness in Tables \ref{tabyields_wind_050}, \ref{tabyields_wind_100}, \ref{tabyields_wind_200} and \ref{tabyields_wind_250} we provide just the wind contribution for these stars.\\

\section{The yields}\label{sec:2}

\begin{figure}
\centering
\includegraphics[width=0.48\textwidth]{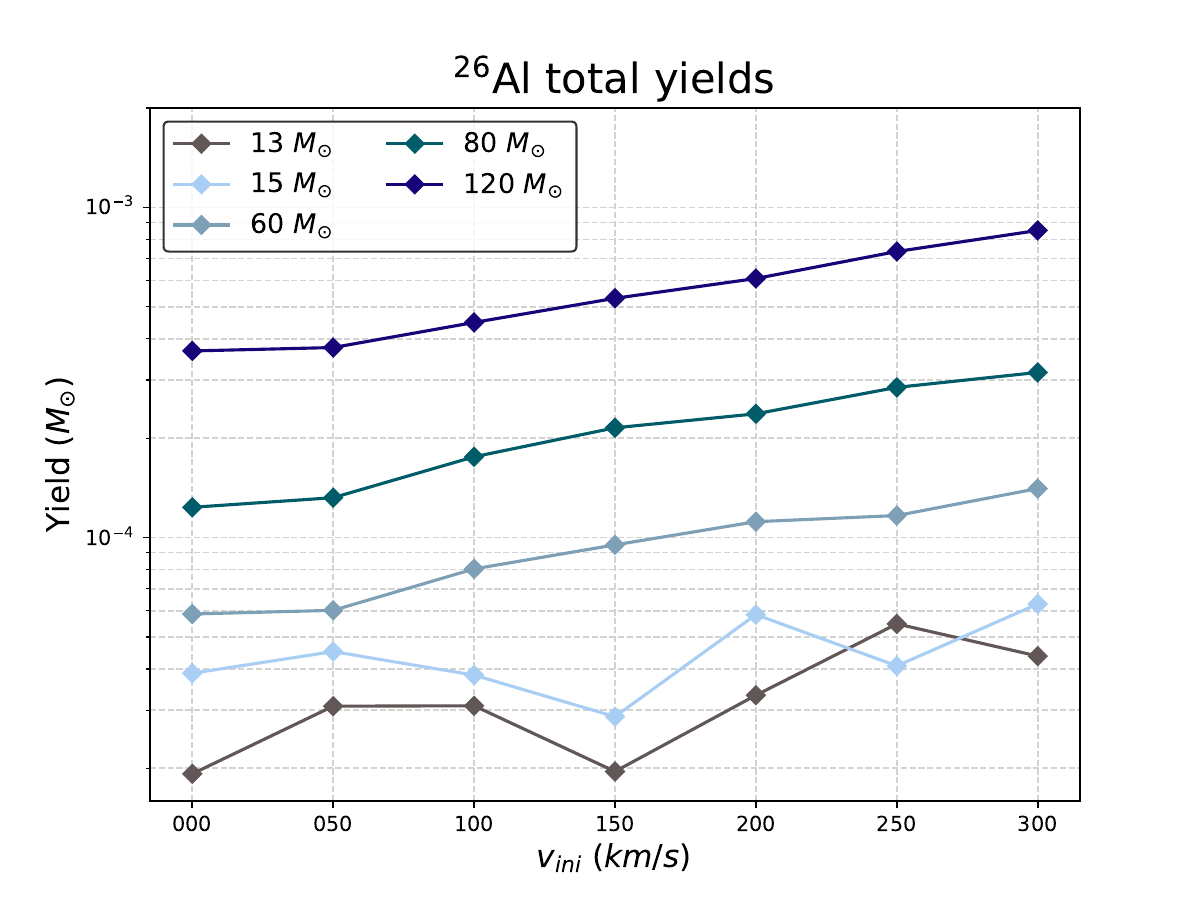}
\caption{\isotope[26]{Al} total yields as a function of the initial rotational velocity, for a selection of masses from the complete set.}
\label{Al26_yields}
\end{figure}

Massive stars synthesize \isotope[26]{Al} in three distinct phases: (1) in central H burning, then spreading it within the H convective core; (2) in the C burning shell, spreading it within the C convective shell; (3) by the explosive C/Ne burning during the explosion. Note that the \isotope[26]{Al} present in the wind is either the one mixed up to the surface by the first dredge-up or the one present beyond the outer border of the He convective shell and that therefore has not been destroyed by the He burning. For a detailed discussion see \LCcite{}. 
Each of the three components scales on average directly with the initial mass. The direct scaling of the \isotope[26]{Al} ejected by the wind with the mass simply reflects the facts that a) the size of the convective core (and hence the buffer within which \isotope[26]{Al} is produced) scales directly with the mass and b) that the mass loss scales directly with the luminosity (and hence the initial mass). Also the overall direct scaling of the yield of \isotope[26]{Al} with the mass in both the C burning shell and the explosion simply reflects the fact that the more massive the star, the more massive both the He and the CO cores, and the more compact the structure at the passage of the shock wave. However, since the advanced burning do not depend just on the CO core mass but are the result of a complex interplay between the mass size of the CO core and the amount of \isotope[12]{C} left by the He burning, the production of \isotope[26]{Al} in both the C convective shell and the explosion does not necessarily scale tightly with the initial mass. Since all three contribution to the synthesis of \isotope[26]{Al} basically scale directly with mass, also the overall yields of \isotope[26]{Al} on average increases with the initial mass.\\
The effect of rotation on this basic scenario is shown in Fig. \ref{Al26_yields}. The figure shows the scaling of the \isotope[26]{Al} yield with the initial rotation velocity for a subsample of our grid of models.
It is quite evident that the yield of \isotope[26]{Al} scales directly with the initial rotation velocity. The explanation of this, as well as the understanding of the not strict direct scaling shown by the stars of lower mass, may be easily understood by reminding that a first important effect of rotation is that of increasing the size of the convective core and of activating a significant mixing of the matter within the whole mass, occurrences that increase the amount of \isotope[26]{Al} produced and preserved in H burning. The second effect of rotation is an increase in the mass loss, occurrence that also helps in increasing the yields of \isotope[26]{Al}. Beyond the He burning, rotation influences the advanced evolutionary phases only indirectly, in the sense that the faster the rotation the more massive the CO core and the lower the amount of \isotope[12]{C} left by the He burning. Both these effect concur to increase the yields of \isotope[26]{Al} even if we do not expect, and actually we do not obtain, a strict direct scaling of the yield with the initial rotation velocity. See the discussion above and \LCcite{}.

\vspace{1cm}
\begin{figure}
\centering
\includegraphics[width=0.48\textwidth]{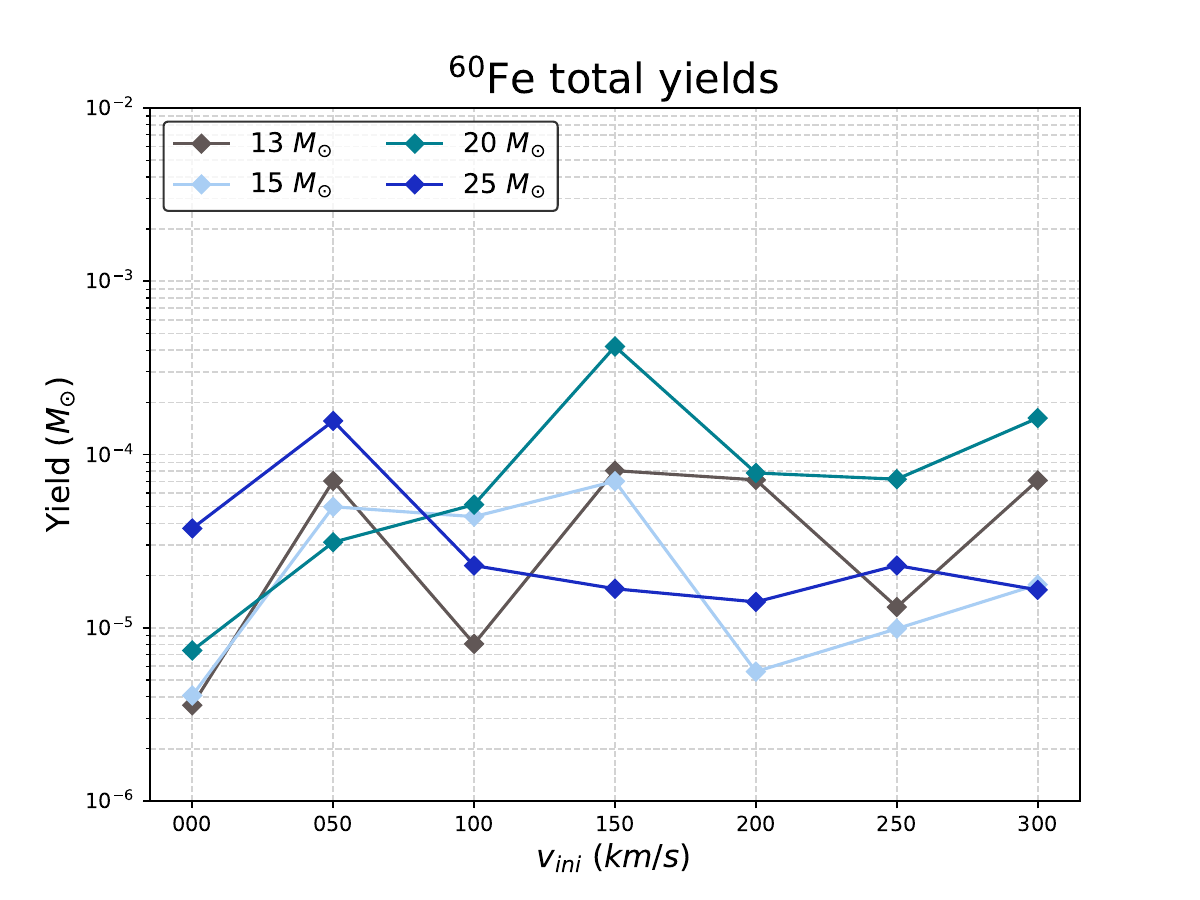}
\caption{\isotope[60]{Fe} total yields as a function of the initial rotational.}
\label{Fe60_yields}
\end{figure}

\isotope[60]{Fe} is produced both in the He and C convective shells, as well as by the explosive burning. Among these three contributions, the yield provided by the C shell dominates in stars with $M < \SI{25}{\smasses}$, while the He shell becomes dominant in more massive stars (see Fig. 6 in \LCcite{} for further details). Since \isotope[60]{Fe} is ejected in the ISM only through the explosion, stars more massive than \SI{25}{\smasses} do not contribute at all to the global inventory of \isotope[60]{Fe} because (in the current scenario) they are assumed to fully collapse in the remnant. Similarly to the \isotope[26]{Al} case, also the yield of \isotope[60]{Fe} shows a quite tight direct scaling with the initial mass in non rotating models (see Fig. 6 in \LCcite{}). The reason is that in this mass range the mass size of the last C convective shell scales with the CO core mass that, in turn, scales directly with the initial mass. Note that, even if the amount of C left by the He burning plays a fundamental role in the formation and extension of the C convective shells, in this case there is a quite tight relation between the amount of C left by the He burning and the CO core mass.  In general, the inclusion of rotation increases \isotope[60]{Fe} yields. This is because, as previously mentioned, rotating models evolve similarly to more massive non-rotating stars for the same initial mass. Thus, they develop fewer C shells due to the lower \isotope[12]{C} fraction left at the end of He burning. However, these shells are more extended, resulting in increased \isotope[60]{Fe} synthesis. Given the non-monotonic nature of this trend, a detailed case-by-case analysis is required to fully understand the impact of initial mass and rotation on \isotope[60]{Fe} production in these stars. That is true in particular for stars with $M < \SI{25}{\smasses}$, in which \isotope[60]{Fe} production happens mainly in the C convective shell. Regarding the $\SI{25}{\smasses}$ models, where \isotope[60]{Fe} production mainly occurs in the He shell, two distinct effects of rotation can be observed. First, rotation-induced mixing during core He burning leads to the formation of a He convective shell in a region with a variable He profile, created by the continuous diffusion of He burning products. This causes the He burning shell to form deeper, reaching higher temperatures even at lower masses, thereby increasing the synthesis of \isotope[60]{Fe}. On the other hand, increased mass loss associated with higher rotational velocities results in less extended He convective shells, reducing \isotope[60]{Fe} production. The first effect dominates on the $v = \SI{50}{\kilo\meter\per\second}$ model, where the yield overcomes that of the non-rotating one. In contrast, the second effect prevails in models with higher initial rotational velocities, leading to a net reduction of the \isotope[60]{Fe} yield.\\

\section{Comparison with the observations}\label{sec:4}

To compare theoretical predictions with observations, the individual stellar yields of \isotope[26]{Al} and \isotope[60]{Fe} must be integrated over a given initial mass function (IMF) to obtain the total contribution from a stellar population

\begin{equation}\label{eq_int_IMF}
Y_i = \int_{M_{bottom}}^{M_{top}}y_{i}(m)m^{-(1+x)}dm\text{ .}
\end{equation}
\quad

Assuming the steady state condition, i.e. the destruction rate equals the production rate, we obtain the \isotope[60]{Fe}/\isotope[26]{Al} $\gamma$-ray flux ratio from the abundance ratio 

\begin{equation}\label{eq_ratio}
\frac{N(^{60}\text{Fe})}{N(^{26}\text{Al})} = \frac{\int_{M_{bottom}}^{M_{top}}\frac{y_{60}(m)}{60}m^{-(1+x)}dm}{\int_{M_{bottom}}^{M_{top}}\frac{y_{26}(m)}{26}m^{-(1+x)}dm}\text{ .}
\end{equation}
\quad

Since we are interested in the galactic production of \isotope[26]{Al} and \isotope[60]{Fe}, we consider the Salpeter IMF, i.e. the one with $x=1.35$, which is assumed to be valid in our Galaxy.\\
In addition, since not all stars rotate at the same velocity, we need to integrate also on a distribution of rotational velocities.
In recent years, telescopes such as the VLT and GAIA allowed high-precision studies of thousands of stellar spectra, providing information on their rotational velocities \citep{Dufton2006, glebocki2000, glebocki2005}. At the same time, Galactic Chemical Evolution (GCE) calculations based on the \LCCCcite{} database were performed, allowing to constrain an empirical Initial Distribution of Rotational Velocities (IDROV) \citep{Prantzos2018} which is able to reproduce most of the abundances in our Galaxy. We decided to test the distributions obtained with both approaches. \\

Firstly, we adopted the Gaussian fit proposed by \citeauthor{Dufton2006} \citeyear{Dufton2006} by means of high-resolution spectroscopy of unevolved targets in two Galactic clusters, NGC 3293 and NGC 4755. They fitted their observed distribution of projected rotational velocities with a Gaussian function, obtaining a maximum at $v = 250k$m/s and $\sigma =  110$km/s. Although they did not measure the initial rotational velocities, the fact that the targets are MS stars allows us to reasonably assume that they have retained most of their initial angular momentum. Hence, by performing a second integration over this gaussian distribution of rotational velocities we obtain a number ratio of $\isotope[60]{Fe}/\isotope[26]{Al} = 0.24$. \citeauthor{Prantzos2018} \citeyear{Prantzos2018}, vice versa, adopted a distribution of \LCCCcite{} rotation velocities tuned in order to reproduce the temporal trend of several elemental species over the lifetime of our galaxy. In particular, in that paper the relative proportion among stars of  0, 150, and 300 km/s was 65:30:5. By adopting a simple linear interpolation among these three velocities we obtain in this case a ratio $\isotope[60]{Fe}/\isotope[26]{Al} = 0.26$.

Note that, though both IDROV give similar $\isotope[60]{Fe}/\isotope[26]{Al}$ ratios, this must be seen as purely accidental. Vice versa it is very interesting to note that the current best estimates of the observed ratio provided by \citeauthor{Wang2020} \citeyear{Wang2020} range between 0.18 and 0.24, values definitely in good agreement with our theoretical prediction.

Interestingly, the $\isotope[60]{Fe}/\isotope[26]{Al}$ ratio provided by just non-rotating models amount to $\num{6.23e-2}$, a value much lower than the one provided by the observations. Hence we conclude that the inclusion of rotation in the evolution of the massive stars is critical if we want to understand the nucleosynthesis of \isotope[26]{Al} and \isotope[60]{Fe}. As a last comment, let us remark that also the non-rotating models published by \LCcite{} provide the same $\isotope[60]{Fe}/\isotope[26]{Al} = \num{6.31e-2}$, if the same explodability range is adopted.

\section{Conclusions}
We have presented a first detailed analysis of the \isotope[26]{Al} and \isotope[60]{Fe} yields produced by a generation of rotating massive stars, based on an extended grid of models, comprising the \LCCCcite{} ones and new computations obtained by means of the same codes and input physics. The explored interval of initial equatorial rotational velocities ranges between 0 and 300 km/s, with a step of 50 km/s.\\

The production of \isotope[26]{Al} shows a quite monotonic trend across the range of masses and rotational velocities. The non-strictly monotonic trend shown by the two less massive stars depends on the fact that the explosive production of \isotope[26]{Al} is the result of a complex interplay between nuclear burning and the extension and timing of the various convective zones. 
\isotope[60]{Fe}, on the contrary, shows a more complex dependence on the initial rotation velocity because its synthesis occurs mainly in the C and He convective shell, whose development, extension and duration are the result of a complex interplay among the mass size of the CO core mass, the amount of C left by the He burning, the rotation-induced instabilities (that lead to an extended mixing of the matter through a large par of the stars) and the strong dependence of mass loss on the luminosity of a star.
Assuming a Salpeter IMF with a slope $x=1.35$, steady state abundances of \isotope[26]{Al} and \isotope[60]{Fe} in the Milky Way and two different distributions of initial equatorial rotation velocity, our models predict an \isotope[60]{Fe}/\isotope[26]{Al} ratios of the order of 0.24. Viceversa, the adoption of just non rotating models leads to a much lower ratio, of the order of 0.06.
Since the latest determination of the observed ratio \citeauthor{Wang2008} \citeyear{Wang2008} ranges between 0.18 and 0.24, we conclude that only the inclusion of rotation in the computation of the yields of these two $\rm \gamma-ray$ emitters provides a satisfactory fit of the currently available observed ratio.

\begin{acknowledgements}
This work has been partially supported by the Theory Grant "Evolution, nucleosynthesis and final fate of stars in the transition between AGB and Massive Stars" (1.05.12.06.04, PI M. Limongi) and by the Theroy Grant “Massive stars as cosmic clocks: shaping the evolution of infant galaxies” (1.05.24.05.07, PI M. Limongi) of the INAF Fundamental Astrophysics Funding Program 2022-2023. LR acknowledges the support from the ChETEC-INFRA -- Transnational Access Projects 22102724-ST and 23103142-ST and the PRIN URKA Grant Number \verb |prin_2022rjlwhn|. ML has also been partially supported by the World Premier International Research Center Initiative (WPI), MEXT, Japan. We acknowledge support from PRIN MUR 2022 (20224MNC5A), "Life, death and after-death of massive stars", funded by European Union - Next Generation EU. This work is the result of a master thesis in the Sapienza University of Rome.
\end{acknowledgements}

\include{tab_050}
\include{tab_100}

\include{tab_200}

\include{tab_250}

\include{tab_wind_050}
\include{tab_wind_100}

\include{tab_wind_200}

\include{tab_wind_250}


\end{document}

%% file: tab_050.tex
\startlongtable
\floattable
\begin{deluxetable}{lccccccccc}
\tabletypesize{\scriptsize}
\tablecolumns{10}
\tablecaption{Isotopic Yields (\SI{}{\smasses}) — 50 km/s. \label{tabyields_050}}
\tablehead{$\rm Isotope$ & $\rm 13~M_\odot$ & $\rm 15~M_\odot$ & $\rm 20~M_\odot$ & $\rm 25~M_\odot$ & $\rm 30~M_\odot$ & $\rm 40~M_\odot$ & $\rm 60~M_\odot$ & $\rm 80~M_\odot$ & $\rm 120~M_\odot$}
\startdata
$\rm^{1}H$ & 5.5167E+00 & 6.1105E+00 & 7.5549E+00 & 9.1644E+00 & 1.0581E+01 & 1.3693E+01 & 1.9709E+01 & 2.4434E+01 & 3.4893E+01\\ 
$\rm^{2}H$ & 2.3053E-16 & 1.7813E-14 & 2.3960E-11 & 5.1510E-10 & 1.3611E-09 & 8.4504E-09 & 2.2671E-07 & 5.0211E-07 & 1.2490E-07\\ 
$\rm^{3}H$ & 4.1594E-25 & 1.4813E-21 & 2.1056E-16 & 4.4193E-14 & 1.9504E-13 & 2.6664E-12 & 1.1256E-11 & 2.8927E-11 & 1.1613E-12\\ 
$\rm^{3}He$ & 3.2284E-04 & 3.4927E-04 & 4.1285E-04 & 4.7453E-04 & 5.2686E-04 & 6.0420E-04 & 9.7693E-04 & 1.1801E-03 & 1.4907E-03\\ 
$\rm^{4}He$ & 4.1115E+00 & 4.6612E+00 & 6.0325E+00 & 7.5683E+00 & 8.5119E+00 & 1.1259E+01 & 2.1402E+01 & 2.9823E+01 & 5.1230E+01\\ 
$\rm^{6}Li$ & 7.6955E-12 & 1.2376E-11 & 2.3654E-11 & 3.0764E-11 & 3.6924E-11 & 4.5097E-11 & 5.7575E-11 & 7.1436E-11 & 3.6865E-11\\ 
$\rm^{7}Li$ & 1.7644E-10 & 1.1382E-09 & 4.9033E-10 & 6.8004E-10 & 8.0723E-10 & 9.7316E-10 & 1.1118E-09 & 1.2941E-09 & 8.6025E-10\\ 
$\rm^{7}Be$ & 3.2620E-13 & 1.6513E-12 & 6.6217E-12 & 7.7653E-12 & 9.3255E-12 & 7.4712E-12 & 2.8615E-11 & 1.9440E-11 & 4.7368E-12\\
\enddata
\end{deluxetable} 

%% file: tab_100.tex
\startlongtable
\floattable
\begin{deluxetable}{lccccccccc}
\tabletypesize{\scriptsize}
\tablecolumns{10}
\tablecaption{Isotopic Yields  (\SI{}{\smasses}) — 100 km/s. \label{tabyields_100}}
\tablehead{$\rm Isotope$ & $\rm 13~M_\odot$ & $\rm 15~M_\odot$ & $\rm 20~M_\odot$ & $\rm 25~M_\odot$ & $\rm 30~M_\odot$ & $\rm 40~M_\odot$ & $\rm 60~M_\odot$ & $\rm 80~M_\odot$ & $\rm 120~M_\odot$}
\startdata
$\rm^{1}H$ & 5.0968E+00 & 5.4737E+00 & 6.6149E+00 & 8.1377E+00 & 9.3062E+00 & 1.2393E+01 & 1.8082E+01 & 2.3326E+01 & 3.3684E+01\\ 
$\rm^{2}H$ & 2.1249E-16 & 1.7151E-14 & 2.5033E-11 & 5.3281E-10 & 1.3679E-09 & 8.4513E-09 & 2.2672E-07 & 5.0005E-07 & 1.2483E-07\\ 
$\rm^{3}H$ & 3.8990E-25 & 1.4813E-21 & 2.1078E-16 & 4.4194E-14 & 1.9505E-13 & 2.6664E-12 & 1.1256E-11 & 2.8922E-11 & 1.1605E-12\\ 
$\rm^{3}He$ & 2.7263E-04 & 2.9176E-04 & 3.2632E-04 & 3.7880E-04 & 4.0519E-04 & 5.0155E-04 & 7.0388E-04 & 9.1821E-04 & 1.3738E-03\\ 
$\rm^{4}He$ & 4.0839E+00 & 4.4143E+00 & 6.2151E+00 & 7.8452E+00 & 9.1885E+00 & 1.4095E+01 & 2.2169E+01 & 3.1002E+01 & 5.5419E+01\\ 
$\rm^{6}Li$ & 4.2572E-12 & 6.8319E-12 & 1.4293E-11 & 2.1334E-11 & 3.1274E-11 & 4.7252E-11 & 6.7630E-11 & 5.7505E-11 & 4.9900E-11\\ 
$\rm^{7}Li$ & 9.4535E-11 & 1.6145E-10 & 3.3772E-10 & 4.5300E-10 & 6.3648E-10 & 8.8707E-10 & 1.5047E-09 & 1.0666E-09 & 1.0571E-09\\ 
$\rm^{7}Be$ & 1.0671E-12 & 7.0233E-12 & 6.3567E-12 & 9.4144E-12 & 1.2233E-11 & 1.3713E-11 & 3.5816E-10 & 1.7250E-10 & 1.1756E-12\\
\enddata
\end{deluxetable}

%% file: tab_200.tex
\startlongtable
\floattable
\begin{deluxetable}{lccccccccc}
\tabletypesize{\scriptsize}
\tablecolumns{10}
\tablecaption{Isotopic Yields  (\SI{}{\smasses}) — 200 km/s. \label{tabyields_200}}
\tablehead{$\rm Isotope$ & $\rm 13~M_\odot$ & $\rm 15~M_\odot$ & $\rm 20~M_\odot$ & $\rm 25~M_\odot$ & $\rm 30~M_\odot$ & $\rm 40~M_\odot$ & $\rm 60~M_\odot$ & $\rm 80~M_\odot$ & $\rm 120~M_\odot$}
\startdata
$\rm^{1}H$ & 4.3574E+00 & 4.9537E+00 & 5.9500E+00 & 7.0191E+00 & 8.0241E+00 & 1.0568E+01 & 1.5657E+01 & 1.9115E+01 & 2.9327E+01\\ 
$\rm^{2}H$ & 1.9275E-16 & 1.5114E-14 & 2.4332E-11 & 5.3299E-10 & 1.4030E-09 & 8.4520E-09 & 2.2672E-07 & 5.0147E-07 & 1.2515E-07\\ 
$\rm^{3}H$ & 4.1599E-25 & 1.4813E-21 & 2.1078E-16 & 4.4192E-14 & 1.9504E-13 & 2.6663E-12 & 1.1256E-11 & 2.8920E-11 & 1.1551E-12\\ 
$\rm^{3}He$ & 1.8573E-04 & 2.0738E-04 & 2.2345E-04 & 2.5586E-04 & 2.6614E-04 & 3.4857E-04 & 4.5862E-04 & 5.0962E-04 & 7.9946E-04\\ 
$\rm^{4}He$ & 4.0553E+00 & 4.4862E+00 & 6.3935E+00 & 8.3844E+00 & 1.0635E+01 & 1.4428E+01 & 2.3309E+01 & 3.2980E+01 & 5.9652E+01\\ 
$\rm^{6}Li$ & 1.6769E-12 & 3.2177E-12 & 9.5175E-12 & 1.3812E-11 & 2.3845E-11 & 3.7186E-11 & 5.9913E-11 & 5.4595E-11 & 4.1453E-11\\ 
$\rm^{7}Li$ & 3.4607E-11 & 7.0809E-11 & 2.1733E-10 & 3.1705E-10 & 5.4030E-10 & 8.1050E-10 & 1.1638E-09 & 1.0857E-09 & 9.1353E-10\\ 
$\rm^{7}Be$ & 3.2479E-12 & 4.0081E-12 & 9.0868E-12 & 1.1581E-11 & 9.1868E-12 & 1.2327E-11 & 1.9785E-11 & 6.9671E-11 & 1.6809E-12\\
\enddata
\end{deluxetable}

%% file: tab_250.tex
\startlongtable
\floattable
\begin{deluxetable}{lccccccccc}
\tabletypesize{\scriptsize}
\tablecolumns{10}
\tablecaption{Isotopic Yields  (\SI{}{\smasses}) — 250 km/s. \label{tabyields_250}}
\tablehead{$\rm Isotope$ & $\rm 13~M_\odot$ & $\rm 15~M_\odot$ & $\rm 20~M_\odot$ & $\rm 25~M_\odot$ & $\rm 30~M_\odot$ & $\rm 40~M_\odot$ & $\rm 60~M_\odot$ & $\rm 80~M_\odot$ & $\rm 120~M_\odot$}
\startdata
$\rm^{1}H$ & 4.1793E+00 & 4.8205E+00 & 5.8563E+00 & 6.6467E+00 & 7.7332E+00 & 9.9696E+00 & 1.3593E+01 & 1.8277E+01 & 2.9079E+01\\ 
$\rm^{2}H$ & 1.8667E-16 & 1.5342E-14 & 2.3621E-11 & 5.2770E-10 & 1.4082E-09 & 8.4528E-09 & 2.2672E-07 & 5.0254E-07 & 1.2522E-07\\ 
$\rm^{3}H$ & 4.1599E-25 & 1.4813E-21 & 2.1078E-16 & 4.4193E-14 & 1.9505E-13 & 2.6665E-12 & 1.1256E-11 & 2.8918E-11 & 1.1520E-12\\ 
$\rm^{3}He$ & 1.5840E-04 & 1.7690E-04 & 1.9953E-04 & 2.1421E-04 & 2.3862E-04 & 3.0370E-04 & 4.1601E-04 & 4.5424E-04 & 6.0675E-04\\ 
$\rm^{4}He$ & 4.1109E+00 & 4.6296E+00 & 6.5872E+00 & 8.6051E+00 & 1.1818E+01 & 1.5165E+01 & 2.5036E+01 & 3.9782E+01 & 6.5511E+01\\ 
$\rm^{6}Li$ & 1.3612E-12 & 2.5789E-12 & 7.8378E-12 & 1.3107E-11 & 2.1412E-11 & 3.2657E-11 & 5.4774E-11 & 5.7866E-11 & 3.7453E-11\\ 
$\rm^{7}Li$ & 2.7876E-11 & 5.5341E-11 & 1.7087E-10 & 2.9261E-10 & 1.3416E-09 & 7.2444E-10 & 1.0849E-09 & 1.2255E-09 & 8.4188E-10\\ 
$\rm^{7}Be$ & 2.7025E-12 & 2.6059E-12 & 4.2403E-12 & 1.0504E-11 & 1.4618E-09 & 4.1434E-11 & 1.8576E-11 & 1.5825E-12 & 2.7253E-12\\
\enddata
\end{deluxetable}

%% file: tab_wind_050.tex
\startlongtable
\floattable
\begin{deluxetable}{lcccc}
\tabletypesize{\scriptsize}
\tablecolumns{5}
\tablecaption{Isotopic Yields (wind contribution) — 50 km/s\label{tabyields_wind_050}}
\tablehead{$\rm Isotope$ & $\rm 13~M_\odot$ & $\rm 15~M_\odot$ & $\rm 20~M_\odot$ & $\rm 25~M_\odot$}
\startdata
$\rm^{1}H$ & 1.5972E+00 & 3.8370E+00 & 7.5549E+00 & 9.1644E+00\\ 
$\rm^{2}H$ & 1.1615E-16 & 1.7753E-14 & 2.3960E-11 & 5.1510E-10\\ 
$\rm^{3}H$ & 4.1594E-25 & 1.4813E-21 & 2.1056E-16 & 4.4193E-14\\ 
$\rm^{3}He$ & 9.8768E-05 & 2.2617E-04 & 4.1285E-04 & 4.7453E-04\\ 
$\rm^{4}He$ & 7.6180E-01 & 1.9246E+00 & 4.6691E+00 & 6.6115E+00\\ 
$\rm^{6}Li$ & 7.6951E-12 & 1.2376E-11 & 2.3654E-11 & 3.0764E-11\\ 
$\rm^{7}Li$ & 1.7515E-10 & 4.5913E-10 & 4.9033E-10 & 6.8004E-10\\ 
$\rm^{7}Be$ & 1.0448E-17 & 1.4922E-14 & 6.6217E-12 & 7.7653E-12\\
\enddata
\end{deluxetable} 

%% file: tab_wind_100.tex
\startlongtable
\floattable
\begin{deluxetable}{lcccc}
\tabletypesize{\scriptsize}
\tablecolumns{5}
\tablecaption{Isotopic Yields (wind contribution) — 100 km/s \label{tabyields_wind_100}}
\tablehead{$\rm Isotope$ & $\rm 13~M_\odot$ & $\rm 15~M_\odot$ & $\rm 20~M_\odot$ & $\rm 25~M_\odot$}
\startdata
$\rm^{1}H$ & 5.0968E+00 & 5.4736E+00 & 6.6149E+00 & 8.1377E+00\\ 
$\rm^{2}H$ & 2.1249E-16 & 1.7151E-14 & 2.5033E-11 & 5.3281E-10\\ 
$\rm^{3}H$ & 3.8990E-25 & 1.4813E-21 & 2.1078E-16 & 4.4194E-14\\ 
$\rm^{3}He$ & 2.7263E-04 & 2.9176E-04 & 3.2632E-04 & 3.7880E-04\\ 
$\rm^{4}He$ & 3.1671E+00 & 3.2016E+00 & 5.4588E+00 & 7.1925E+00\\ 
$\rm^{6}Li$ & 4.2572E-12 & 6.8319E-12 & 1.4293E-11 & 2.1334E-11\\ 
$\rm^{7}Li$ & 9.4535E-11 & 1.6145E-10 & 3.3772E-10 & 4.5300E-10\\ 
$\rm^{7}Be$ & 1.0671E-12 & 7.0227E-12 & 6.3567E-12 & 9.4144E-12\\
\enddata
\end{deluxetable}

%% file: tab_wind_200.tex
\startlongtable
\floattable
\begin{deluxetable}{lcccc}
\tabletypesize{\scriptsize}
\tablecolumns{5}
\tablecaption{Isotopic Yields (wind contribution) — 200 km/s\label{tabyields_wind_200}}
\tablehead{$\rm Isotope$ & $\rm 13~M_\odot$ & $\rm 15~M_\odot$ & $\rm 20~M_\odot$ & $\rm 25~M_\odot$}
\startdata
$\rm^{1}H$ & 4.3574E+00 & 4.9537E+00 & 5.9500E+00 & 7.0191E+00\\ 
$\rm^{2}H$ & 1.9275E-16 & 1.5114E-14 & 2.4332E-11 & 5.3299E-10\\ 
$\rm^{3}H$ & 4.1599E-25 & 1.4813E-21 & 2.1078E-16 & 4.4192E-14\\ 
$\rm^{3}He$ & 1.8573E-04 & 2.0738E-04 & 2.2345E-04 & 2.5586E-04\\ 
$\rm^{4}He$ & 3.2835E+00 & 3.7478E+00 & 5.8753E+00 & 7.8572E+00\\ 
$\rm^{6}Li$ & 1.6769E-12 & 3.2177E-12 & 9.5175E-12 & 1.3812E-11\\ 
$\rm^{7}Li$ & 3.4607E-11 & 7.0809E-11 & 2.1733E-10 & 3.1705E-10\\ 
$\rm^{7}Be$ & 3.2479E-12 & 4.0081E-12 & 9.0868E-12 & 1.1581E-11\\
\enddata
\end{deluxetable}

%% file: tab_wind_250.tex
\startlongtable
\floattable
\begin{deluxetable}{lcccc}
\tabletypesize{\scriptsize}
\tablecolumns{5}
\tablecaption{Isotopic Yields (wind contribution) — 250 km/s\label{tabyields_wind_250}}
\tablehead{$\rm Isotope$ & $\rm 13~M_\odot$ & $\rm 15~M_\odot$ & $\rm 20~M_\odot$ & $\rm 25~M_\odot$}
\startdata
$\rm^{1}H$ & 4.1793E+00 & 4.8205E+00 & 5.8563E+00 & 6.6467E+00\\ 
$\rm^{2}H$ & 1.8667E-16 & 1.5342E-14 & 2.3621E-11 & 5.2770E-10\\ 
$\rm^{3}H$ & 4.1599E-25 & 1.4813E-21 & 2.1078E-16 & 4.4193E-14\\ 
$\rm^{3}He$ & 1.5840E-04 & 1.7690E-04 & 1.9953E-04 & 2.1421E-04\\ 
$\rm^{4}He$ & 3.4385E+00 & 4.0494E+00 & 6.1002E+00 & 8.1140E+00\\ 
$\rm^{6}Li$ & 1.3612E-12 & 2.5789E-12 & 7.8378E-12 & 1.3107E-11\\ 
$\rm^{7}Li$ & 2.7876E-11 & 5.5341E-11 & 1.7087E-10 & 2.9261E-10\\ 
$\rm^{7}Be$ & 2.7025E-12 & 2.6059E-12 & 4.2403E-12 & 1.0504E-11\\
\enddata
\end{deluxetable}